\documentclass[prb,twocolumn,superscriptaddress,aps,amssymb,footinbib]{revtex4} 
\usepackage{amssymb}
\usepackage{graphicx}
\usepackage{amsmath}
\usepackage{times}
\usepackage{color}

\def\Tr{{\rm\,Tr}}

\begin{document}

\title{Energy spread and current-current correlation in quantum systems}
\date{\today}

\author{Yonghong Yan }
\email[Email: ] {yhyan@usx.edu.cn}
 \affiliation{Department of
Physics, Shaoxing University, Shaoxing 312000, China}

\author{Feng Jiang}
\email[Email: ] {fjiang@shiep.edu.cn}
 \affiliation{School of Mathematics and Physics, Shanghai University of Electric
Power, Shanghai 200090, China}

\author{Hui Zhao}
 \affiliation{Key Laboratory for Advanced Microstructure Materials of the Ministry
of Education and Department of Physics, Tongji University, Shanghai
200092, China}

\begin{abstract}
We consider energy (heat) transport in quantum systems, and
establish a relationship between energy spread and energy
current-current correlation function. The energy current-current
correlation is related to thermal conductivity by the Green-Kubo
formula, and thus this relationship allows us to study conductivity
directly from the energy spread process. As an example, we
investigate a spinless fermion model; the numerical results confirm
the relationship.
\end{abstract}

\maketitle
\section{Introduction}
Heat transport has attracted increasing interest recently in both
classical nonlinear lattices and quantum
systems\cite{NLi12,Lepri,Dhar,Liu13,Zotos}. In this field, a
particularly interesting problem is related to the issue of
Fourier's law. Considering for example the normal transport in
one-dimensional systems, Fourier's law states that
$j(x,t)=-\kappa\partial_xT(x,t)$, where $j(x,t)$ is the local heat
current, $T(x,t)$ is the local equilibrium temperature, and $\kappa$
is the heat conductivity. If we let $\varepsilon(x,t)$ denote the
local energy density, then the continuity equation reads
$\partial_t\varepsilon(x,t)+\partial_xj(x,t)=0$. Combining these
equations with $\frac{\partial\varepsilon}{\partial T}=c$, where $c$
is the specific heat per unit volume, we arrive at the energy
diffusion equation,
$\partial_t\varepsilon(x,t)=\frac{\kappa}{c}\partial^2_x\varepsilon(x,t)$.

In classical systems, it was shown that normal diffusion can be
characterized by the mean squared displacement of the Helfand moment
\cite{Helfand,Viscardy}, which is related to the autocorrelation
function of heat current and thus to the Green-Kubo formula. Beyond
the normal diffusion, some recent works have investigated the
relation between heat diffusion and conduction
\cite{Denisov,Li03,Zhao,Zaburdaev,Dhar13,YLi}. In particular, a
rigorous relationship  between energy (heat) spread and  heat
conduction has been established from statistical
principles\cite{Liu14}. Therein, an excess energy distribution is
introduced, and then the energy diffusion is characterized by the
mean square deviation (MSD) of energy, which is connected to the
autocorrelation function of heat current. Accordingly, how thermal
conductivity depends on the system size may be extracted from energy
diffusion in lattice systems \cite{Zhao,Cipriani,NLi}.

Simultaneously, heat transport in low-dimensional quantum systems
has also been investigated intensively
\cite{Zotos96,Zotos99,Narozhny,Meisner,Benz,Karrasch,Zotos04,Jung,Karrasch13,Znidaric}.
A commonly used method is the Green-Kubo formula within linear
response theory, where nonzero Drude weights usually indicate
ballistic transport. An interesting example is the ballistic energy
transport in the spin-$\frac{1}{2}$ XXZ chain due to the
conservation of the current operator
\cite{Zotos97,Klumper,Karrasch14}. Besides, quantum quench dynamics
or spreading of different densities (e.g., energy densities) has
also been studied\cite{Michel,Langer,Steinigeweg09}. To determine
whether the spread process is  ballistical, diffusive, or of other
type, one can observe the time evolution of the spatial variance
$\sigma^2$ (or MSD) of certain nonequilibrium density. For ballistic
transport, the variance behaviors as  $\sigma^2\sim t^2$ whereas for
diffusive transport  $\sigma^2\sim t$. In Ref.
\onlinecite{Steinigeweg09}, a connection between the variance and
the current-autocorrelation function  was proposed; however, it is
applicable at high temperatures. The general connection between the
spreading processes and transport properties such as heat
conductivity is not well understood yet.

In this paper, starting from an energy density distribution, we give
a general connection between the MSD of energy diffusion and the
autocorrelation function of energy current for quantum systems,
within the linear response theory. This offers a different way to
extract thermal  conductivity from the energy spreading process. As
an example, we apply it to a spinless fermion model, and the
numerical results confirm this connection.

\section{Connection between MSD and current-autocorrelation function}
In the following, we restrict to the one-dimensional case. The
generalization to higher-dimensional systems is straightforward. The
system is typically described by a continuous Hamiltonian:
\begin{eqnarray}
H_0=\int h(x)dx.
\end{eqnarray}
At the infinite past an  additional perturbation, $H'=-\int
\eta(x)h(x)dx$, is also applied to the system. Here $\eta(x)$ is
nonzero only in a local region. Thus the total Hamiltonian reads
$H=H_0+H'$. Before $t=0$, we suppose the system is described by a
canonical ensemble at temperature $T$. Then the partition function
is $Z=\Tr( e^{-\beta H})$, where $\beta=1/k_BT$. At time $t=0$, the
perturbation is turned off suddenly. After that the quenched initial
nonequilibrium state begins to relax towards the equilibrium state,
and so does the local energy distribution. The local excess energy
at $t>0$ can be described by
\begin{eqnarray}
\delta\langle h(x,t)\rangle_{neq} \equiv \langle h(x,t)\rangle_{neq}
-\langle h(x)\rangle, \label{Eq_deltaH}
\end{eqnarray}
where $h(x,t)=e^{iH_0t/\hbar}h(x)e^{-iH_0t/\hbar}$. $\langle \cdot
\rangle_{neq}$ denotes the expectation value in the nonequilibrium
state, i.e., $\langle \cdot \rangle_{neq}=\Tr(e^{-\beta H}\cdot)/Z$,
and $\langle \cdot \rangle$ denotes the equilibrium average,
$\langle \cdot \rangle=\Tr(\rho_0\cdot)$ with $\rho_0=e^{-\beta
H_0}/\Tr(e^{-\beta H_0})$.

To evaluate $\langle h(x,t)\rangle_{neq}$,
 we consider  an operator
$U(\tau)=e^{H_0\tau/\hbar}e^{-H\tau/\hbar}$.  The equation of motion
of $U(\tau)$ is
\begin{eqnarray}
-\hbar\frac{\partial U(\tau)}{\partial\tau}=H'(\tau)U(\tau),
\end{eqnarray}
where $H'(\tau)=e^{H_0\tau/\hbar}H'e^{-H_0\tau/\hbar}$. To the first
order of $H'$, the solution can be written as
$U(\tau)\approx1-\frac{1}{\hbar}\int_0^{\tau}d\tau'H'(\tau')$. Thus
we may have
\begin{eqnarray}
e^{-\beta H}&=&e^{-\beta H_0}U(\hbar\beta) \nonumber \\
&\approx&e^{-\beta H_0}-\frac{1}{\hbar}e^{-\beta
H_0}\int_0^{\hbar\beta}d\tau H'(\tau). \label{Eq_betaH}
\end{eqnarray}
Then we can obtain the partition function to the first order of
$H'$,
\begin{eqnarray}
Z/Z_0\approx1-\frac{1}{\hbar}\langle\int_0^{\hbar\beta}d\tau
H'(\tau)\rangle.\label{Eq_Z}
\end{eqnarray}
Substituting Eqs.~(\ref{Eq_betaH}) and (\ref{Eq_Z}) into
Eq.~(\ref{Eq_deltaH}),  we can obtain after some algebra,
\begin{eqnarray}
\delta\langle h(x,t)\rangle_{neq}
&=&-\frac{1}{\hbar}\langle\int_0^{\hbar\beta}d\tau
H'(\tau)h(x,t)\rangle \nonumber \\
& &+\frac{1}{\hbar}\langle\int_0^{\hbar\beta}d\tau H'(\tau)\rangle
\langle h(x,t)\rangle. \label{delta_h}
\end{eqnarray}
The second term in Eq.~(\ref{delta_h}) is time-independent actually.
The probability distribution function is then defined as
\begin{eqnarray}
\rho_E(x,t)=\delta\langle h(x,t)\rangle_{neq}/\mathcal{N},
\end{eqnarray}
where $\mathcal{N}=\int dx \delta\langle h(x,t)\rangle_{neq}=Tc\int
dx'\eta(x')$ is a normalization constant; see appendix
\ref{appendixA}.

The mean square deviation for energy spread is then
\begin{eqnarray}
\langle \Delta x^2(t)\rangle_E &\equiv&\int(x-\langle
x\rangle_E)^2\rho_E(x,t)dx \nonumber \\
&=&\langle x^2(t)\rangle_E-\langle x\rangle^2_E.
\end{eqnarray}
Using the reasoning similar to Ref. \onlinecite{Liu14}, it can be
shown that $\langle x\rangle_E$ is a constant. For later
convenience, we introduce  two correlation functions:
\begin{eqnarray}
C_{jj}(x't',xt)&=&\langle\int_0^{\hbar\beta}d\tau
j(x',t'-i\tau)j(x,t)\rangle
\end{eqnarray}
and
\begin{eqnarray}
C_{hh}(x't',xt)&=&\langle\int_0^{\hbar\beta}d\tau
h(x',t'-i\tau)h(x,t)\rangle,
\end{eqnarray}
where the current operator $j(x,t)$ is defined via the continuity
equation, $\partial_t h(x,t)+\partial_x j(x,t)=0$. For homogeneous
systems, these correlation functions are invariant under both
temporal  translation and  spacial translation, i.e.,
$C_{jj}(x't',xt)=C_{jj}(x-x',t-t')$; a similar relation holds for
$C_{hh}$. Further, we can have
$\partial^2_tC_{hh}(x't',xt)=\partial^2_xC_{jj}(x't',xt)$\cite{Liu14}.

Making use of the above equations, we can obtain
\begin{eqnarray}
\mathcal{N}\frac{d^2\langle x^2(t)\rangle_E}{dt^2}
&=&\frac{1}{\hbar}\int\! dx\int\! dx' x^2
\frac{d^2C_{jj}(x'0,xt)}{dx^2}\eta(x') \nonumber \\
&=&\frac{1}{\hbar}\int\! dx x^2 \frac{d^2C_{jj}(x,t)}{dx^2}\!\int\!
dx'\eta(x').
\end{eqnarray}
That means
\begin{eqnarray}
\frac{d^2\langle x^2(t)\rangle_E}{dt^2} =\frac{1}{\hbar Tc}\int\! dx
x^2 \frac{d^2C_{jj}(x,t)}{dx^2}. \label{connect}
\end{eqnarray}
Eq.~(\ref{connect}) is a rigorous result. Integrating by parts twice
and neglecting the boundary terms, we obtain the final result:
\begin{eqnarray}
\frac{d^2\langle x^2(t)\rangle_E}{dt^2}=\frac{2C_{JJ}(t)}{cT},
\label{central_eq}
\end{eqnarray}
where $ C_{JJ}(t)=\lim_{L\rightarrow \infty}\frac{1}{L}
\int_0^{\beta}d\lambda\langle J(-i\lambda\hbar)J(t)\rangle$ is the
current-current correlation function that appears in the Green-Kubo
formula for heat conductivity; see appendix \ref{appendixB}. Here
$J=\int dx j(x)$ and $L$ is the length of the system. It should be
pointed out that taking the limit $L\rightarrow \infty$ is
necessary, because in systems with a finite size the autocorrelation
function $C_{jj}(x,t)$ at low temperatures may not decay to zero as
$x$ approaches to boundaries (see the example below). In that case,
the boundary terms such as $C_{jj}(x,t)x|^{L/2}_{-L/2}$ need to be
taken into account explicitly.

It is straightforward to extend Eq.~(\ref{central_eq}) to other
conserved quantities of the form $\hat{Q}=\int dx \hat{q}(x)$,
\begin{eqnarray}
\frac{d^2\langle x^2(t)\rangle_Q}{dt^2}=\frac{2C_{J_qJ_q}(t)}{\beta
\sigma^2_Q}, \label{central_eq2}
\end{eqnarray}
where $\langle x^2(t)\rangle_Q$ is defined through the distribution
$\delta\langle \hat{q}(x,t)\rangle_{neq}$, and $\sigma^2_Q=(\langle
\hat{Q}^2\rangle-\langle \hat{Q}\rangle^2)/L$ is the fluctuation of
quantity $Q$. The total current $\hat{J}_q$ is given by
$\hat{J}_q=\int dx \hat{j}_q(x)$, and $\hat{j}_q(x)$ is defined via
the continuity equation $\partial_t \hat{q}(x,t)+\partial_x
\hat{j}_q(x,t)=0$. According to the time evolution of $\langle
x^2(t)\rangle_Q$, transport processes may be classified as diffusive
($\langle x^2(t)\rangle_Q\sim t^{\beta}$, $\beta=1$),
super-diffusive ($1<\beta\leq2$), and  sub-diffusive ($0<\beta<1$).

\section{An example: spinless fermion model}

\begin{figure}
\centering
\includegraphics[width=0.8\columnwidth]{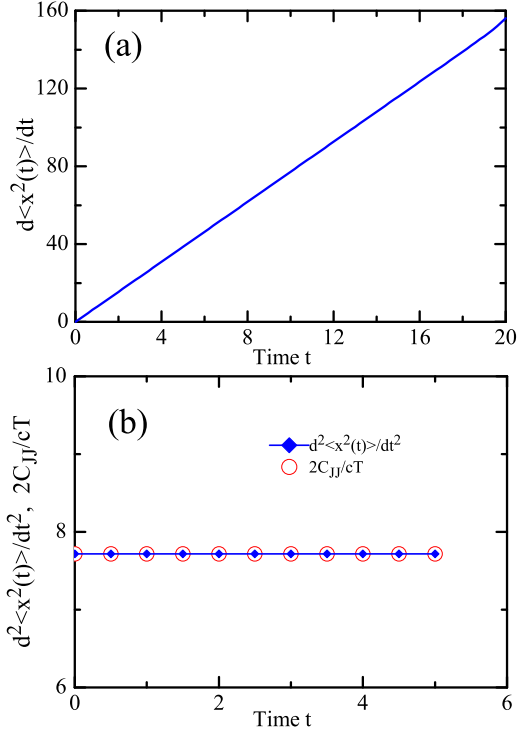}
\caption{(Color online) (a) The first derivative of the MSD with
respect to time $d\langle x^2(t)\rangle_E/dt$  as a function of
time. (b) Comparison of $\frac{d^2\langle x^2(t)\rangle_E}{dt^2}$
and $2C_{JJ}(t)/cT$. The parameters are $t_0=1$ and $T=0.1$, and we
choose $k_B=\hbar=1$. The system size is $L=100$. } \label{fig1}
\end{figure}


As an application of Eq.~(\ref{central_eq}), we consider a
noninteracting fermion model, which may also be viewed as a
spin-$\frac{1}{2}$ XY chain. For complicated systems, one may resort
to methods such as finite-temperature, real-time density matrix
renormalization group \cite{Karrasch}.  The Hamiltonian we consider
reads
\begin{eqnarray}
H_0=-t_0\sum_{i=-L/2+1}^{L/2} (c^{\dagger}_ic_{i+1}+\mathrm{H.c.})
\equiv \sum_{i}h_i,
\end{eqnarray}
where $h_i$ is the local energy operator. The total size of the
system is $L$, and we adopt periodic boundary conditions.  Via the
continuity equation, the current operator can be shown to be
$j_i=\frac{it_0^2}{\hbar}(c^{\dagger}_{i-1}c_{i+1}-c^{\dagger}_{i+1}c_{i-1})$.
The total energy current operator $J=\sum_i j_i$ commutes with
$H_0$, and thus is conserved.  Note that usually energy current is
different from heat current \cite{Mahan}. However in the following
we set the chemical potential to zero, so these two currents are the
same in our case.

We assume a local perturbation, $H'=-\sum_i\eta_ih_i$, where
$\eta_i=0.2$ for $i=0$ and $\eta_i=0$ otherwise. To compute the MSD
of energy diffusion, we first evaluate Eq.~(\ref{delta_h}). In the
basis of single-particle eigenstates of $H_0$
($H_0|\alpha\rangle=\epsilon_{\alpha}|\alpha\rangle$), the operators
can be expressed  as
\begin{eqnarray}
h_i(t)=\sum_{\alpha\beta}\langle \alpha|h_i|\beta\rangle
c^{\dagger}_{\alpha}c_{\beta}e^{i(\epsilon_{\alpha}-\epsilon_{\beta})t/\hbar}
\label{exp1}
\end{eqnarray}
and
\begin{eqnarray}
H'(\tau)=\sum_{\alpha\beta}\langle \alpha|H'|\beta\rangle
c^{\dagger}_{\alpha}c_{\beta}e^{(\epsilon_{\alpha}-\epsilon_{\beta})\tau/\hbar},\label{exp2}
\end{eqnarray}
where $c^{\dagger}_{\alpha}$ ($c_{\alpha}$) creates (destroys) a
particle occupying the state $|\alpha\rangle$. Substituting
Eqs.~(\ref{exp1}) and  (\ref{exp2}) into Eq.~(\ref{delta_h}), we can
get after some algebra
\begin{eqnarray}
\delta\langle h_i(t)\rangle_{neq} =\sum_{\alpha\beta}\langle
\beta|H'|\alpha\rangle\langle \alpha|h_i|\beta\rangle
e^{i(\epsilon_{\alpha}-\epsilon_{\beta})t/\hbar}
\frac{f_{\alpha}-f_{\beta}}{\epsilon_{\alpha}-\epsilon_{\beta}}, \nonumber  \\
\label{delta_hi}
\end{eqnarray}
where $f_{\alpha}=1/(1+e^{\epsilon_{\alpha}/k_BT})$ is the
Fermi-Dirac distribution with the chemical potential being zero and
we have used the identity
$\Tr[\rho_0c^{\dagger}_{\beta}c_{\alpha}c^{\dagger}_{\gamma}c_{\delta}]
=\delta_{\alpha \gamma}\delta_{\beta
\delta}f_{\beta}(1-f_{\alpha})+\delta_{\alpha \beta}\delta_{\gamma
\delta}f_{\beta}f_{\gamma}$. The specific heat can be easily
evaluated from $C_V=\frac{\partial E}{\partial T}$, where
$E=\sum_{\alpha}\epsilon_{\alpha}f_{\alpha}$. In a similar way we
can evaluate $C_{JJ}(t)$, and the final result is
\begin{eqnarray}
C_{JJ}(t)&=&-\frac{1}{L}\sum_{\alpha \beta}\langle
\beta|\hat{J}|\alpha\rangle\langle \alpha|\hat{J}|\beta\rangle
\nonumber \\
& &
\times\frac{f_{\alpha}-f_{\beta}}{\epsilon_{\alpha}-\epsilon_{\beta}}
e^{i(\epsilon_{\alpha}-\epsilon_{\beta})t/\hbar}.
\end{eqnarray}

\begin{figure}
\centering
\includegraphics[width=0.98\columnwidth]{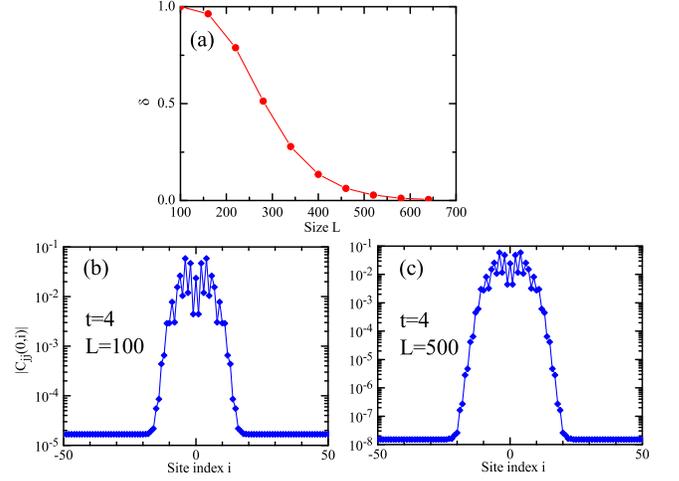}
\caption{(Color online) (a) Difference between the MSD of energy
diffusion and $2C_{JJ}(t)/cT$ as a function of the system size. (b)
and (c) $|C_{jj}(0,i)|$  in logarithm scale at time $t=4$. In (b)
the total size is $L=100$ whereas in (c) $L=500$.  The temperature
is $T=0.01$. $k_B=\hbar=1$. } \label{fig2}
\end{figure}

In numerical simulations, we take $t_0=1$ as units of energy, and we
set $k_B=\hbar=1$. In Fig.~\ref{fig1}(a), we plot the first
derivative of $\langle x^2(t)\rangle_E$ with respect to time at a
high temperature $T=0.1$. The corresponding  real  temperature is of
order $10^3$ K, and a finite size $L=100$ is used here. We see that
$d\langle x^2(t)\rangle_E/dt$ linearly increases with time. This is
also reflected in Fig.~\ref{fig1}(b), where $\frac{d^2\langle
x^2(t)\rangle_E}{dt^2}$ is clearly a constant. Thus the transport
process is  ballistic. To check the validity of
Eq.~(\ref{central_eq}), we plot $\frac{d^2\langle
x^2(t)\rangle_E}{dt^2}$ and $2C_{JJ}(t)/cT$ as functions of time in
Fig.~\ref{fig1}(b), and a good agreement can be observed.

At low temperatures, the finite size effect becomes prominent; i.e.,
there could be a big difference between $\frac{d^2\langle
x^2(t)\rangle_E}{dt^2}$ and $2C_{JJ}(t)/cT$ when the system size is
not large enough. To characterize this difference, we plot the
relative error
\begin{eqnarray}
\delta=|\frac{d^2\langle
x^2(t)\rangle_E}{dt^2}-\frac{2C_{JJ}(t)}{cT}|/|\frac{d^2\langle
x^2(t)\rangle_E}{dt^2}|
\end{eqnarray}
in Fig.~\ref{fig2}(a). As the size increases, the error $\delta$
decays to zero rapidly. The reason for the big difference at a small
size is that the current autocorrelation function $C_{jj}(x'=0,x=i)$
does not decay to zero at the boundaries. For $L=100$, $C_{jj}(0,i)$
takes an appreciably small value at the boundaries, while
$C_{jj}(0,i)$ becomes very small at the boundaries for $L=500$; see
Figs.~\ref{fig2}(b) and (c). Thus when integrating
Eq.~(\ref{connect}) to get Eq.~(\ref{central_eq}), we can not
neglect the boundary terms for small systems. When the boundary
terms are taken into account, we have found excellent agreement
between $\frac{d^2\langle x^2(t)\rangle_E}{dt^2}$ and
$2C_{JJ}(t)/cT$ plus boundary terms regardless of the system size.

\section{Conclusions}
In summary, within the linear response theory we have established a
connection between the MSD of energy diffusion and the
autocorrelation function of energy current for quantum systems,
i.e., $\frac{d^2\langle
x^2(t)\rangle_E}{dt^2}=\frac{2C_{JJ}(t)}{cT}$. It is straightforward
to extend it to other conserved quantities. As an example, we have
applied it to a spinless fermion model (or the spin-$1/2$ XY model).
We found that at high temperatures $\frac{d^2\langle
x^2(t)\rangle_E}{dt^2}$ is consistent with $2C_{JJ}(t)/cT$ even for
a comparatively small size $L=100$. However, at low temperatures,
there may be large difference between $\frac{d^2\langle
x^2(t)\rangle_E}{dt^2}$ and $2C_{JJ}(t)/cT$ when the system
 size  is small due to the ignorance of boundary terms. Indeed when the
boundary terms are included, we could still find excellent agreement
between $\frac{d^2\langle x^2(t)\rangle_E}{dt^2}$ and
$2C_{JJ}(t)/cT$ plus boundary terms regardless of the system size.
This connection thus offers an alternative way to extract
conductivity from the energy spreading process in quantum systems.

\begin{acknowledgments}
The work was supported by NSFC (Grant No. 11204180).
\end{acknowledgments}

\appendix
\section{Normalization constant}
\label{appendixA} Here we will show $\mathcal{N}=\int dx
\delta\langle h(x,t)\rangle_{neq}=Tc\int dx'\eta(x')$. From
Eq.~(\ref{delta_h}), we see that $\mathcal{N}$ consists of two
terms. The first term is
\begin{eqnarray}
& &-\frac{1}{\hbar}\int dx\langle\int_0^{\hbar\beta}d\tau
H'(\tau)h(x,t)\rangle \nonumber \\
&=&\frac{1}{\hbar}\int dx\int dx'\langle\int_0^{\hbar\beta}d\tau
h(x',\tau)h(x,t)\rangle \eta(x') \nonumber \\
&=&\frac{1}{\hbar}\langle\int_0^{\hbar\beta}d\tau
h(0,\tau)H_0\rangle \int dx'\eta(x') \nonumber \\
&=&\frac{1}{\hbar L}\int dy\langle\int_0^{\hbar\beta}d\tau
h(y,\tau)H_0\rangle \int dx'\eta(x') \nonumber \\
&=&\frac{\beta}{L}\langle H_0^2\rangle \int dx'\eta(x'),
\end{eqnarray}
where we have used the spacial-translation invariance of
$C_{hh}(x'0,xt)$. The second term is
\begin{eqnarray}
& &\frac{1}{\hbar}\langle\int_0^{\hbar\beta}d\tau H'(\tau)\rangle
\langle h(x,t)\rangle \nonumber \\
&=& -\frac{1}{\hbar}\int dx' \langle\int_0^{\hbar\beta}d\tau
h(x',\tau)\rangle \eta(x') \langle H_0\rangle \nonumber \\
&=& -\frac{1}{\hbar} \langle\int_0^{\hbar\beta}d\tau
h(0,\tau)\rangle \langle H_0\rangle \int dx' \eta(x') \nonumber \\
&=& -\frac{1}{\hbar L} \int dy\langle\int_0^{\hbar\beta}d\tau
h(y,\tau)\rangle \langle H_0\rangle \int dx' \eta(x') \nonumber \\
&=& -\frac{\beta}{L} \langle H_0\rangle^2 \int dx' \eta(x').
\end{eqnarray}

Upon combining these two terms, we have
$\mathcal{N}=\frac{\beta}{L}[\langle H_0^2\rangle-\langle
H_0\rangle^2]\int dx' \eta(x')$. From $C_V=\frac{\partial
E}{\partial T}$, where $E=\Tr(e^{-\beta H_0}H_0)/\Tr(e^{-\beta
H_0})$, we can obtain $\langle H_0^2\rangle-\langle
H_0\rangle^2=TC_V/\beta$. So we have
\begin{eqnarray}
\mathcal{N}=Tc\int dx'\eta(x').
\end{eqnarray}

\section{Green-Kubo formula for heat conductivity}
\label{appendixB} Here we give a very brief introduction to the
Green-Kubo formula. We begin with the following partition function,
\begin{equation}
Z_0=\Tr[e^{-\beta H_0}].
\end{equation}
Applying a temperature gradient across the system [$T(x)=T+\delta
T(x)$] and assuming local equilibrium, then we may expect
\begin{eqnarray}
\beta H_0&\rightarrow&  \beta\int dxh(x)[1-\frac{\delta T(x)}{T}]
\nonumber \\
 &=&\beta[\int dxh(x)-\int dx h(x)\frac{\delta T(x)}{T}].
\end{eqnarray}
We can treat the second term as a perturbation
\begin{eqnarray}
 H'=-\int dx h(x)\frac{\delta T(x)}{T}\equiv \hat{F}.
\end{eqnarray}
Then within the linear response theory the heat current can be
written as\cite{Mahan}
\begin{eqnarray}
j(x)&=&\int^0_{-\infty}dte^{\eta t}\int_0^{\beta}d\lambda \nonumber \\
& &\times\mathrm{Tr}[\rho_0  \int dx'
 j(x',t-i\lambda\hbar)j(x)\frac{1}{T}\frac{\partial \delta T}{\partial
 x'}],
\end{eqnarray}
where $\eta=0^+$ and $\rho_0=e^{-\beta H_0}/Z_0$. Assuming a uniform
temperature gradient [$\frac{\partial \delta T}{\partial
 x'}=\mathrm{const}.$],
we thus obtain heat conductivity:
\begin{eqnarray}
\kappa&=&\mathrm{Re}\left\{\frac{1}{LT}\int_0^{\infty}dte^{-\eta
t}\int_0^{\beta}d\lambda \mathrm{Tr}[\rho_0
 J(-t-i\lambda\hbar)J]\right\}  \nonumber \\
 &=&\mathrm{Re}\left\{\frac{1}{T}\int_0^{\infty}dt e^{-\eta
t} C_{JJ}(t)\right\},
\end{eqnarray}
where $J=\int dx j(x)$ and
 \begin{eqnarray}
 C_{JJ}(t)=\lim_{L\rightarrow \infty}\frac{1}{L}
 \langle \int_0^{\beta}d\lambda J(-i\lambda\hbar)J(t) \rangle .
\end{eqnarray}
 The above equation may also be cast in
the following form
\begin{eqnarray}
\kappa=\frac{1}{Lk_BT^2}\frac{1}{2}\int_0^{\infty}dt \langle
\{J,J(t)\}\rangle,
\end{eqnarray}
which can be shown in the basis of eigenstates of $H_0$. Although
the Green-Kubo formula for heat conductivity is questionable, we
will not go into the details here \cite{Dhar,Gemmer06}.

\end{document}